\UseRawInputEncoding

\documentclass[10pt, prb, aps, twocolumn, showpacs, citeautoscript, floatfix, reprint, amsmath, amssymb, notitlepage, superscriptaddress]{revtex4-1}

\usepackage{graphicx}
\usepackage{stmaryrd}
\usepackage{rotating}
\usepackage{amsmath}
\usepackage{amsfonts}
\usepackage{amssymb}
\usepackage[countmax]{subfloat}
\usepackage{dcolumn} 
\usepackage{bm} 
\usepackage{color}

\usepackage{float}
\usepackage{latexsym}

\begin{document}

\title{Mapping quantum geometry and quantum phase transitions to real space
\\
by a fidelity marker}




\author{Matheus S. M. de Sousa}

\affiliation{Department of Physics, PUC-Rio, 22451-900 Rio de Janeiro, Brazil}

\author{Antonio L. Cruz}

\affiliation{Department of Physics, PUC-Rio, 22451-900 Rio de Janeiro, Brazil}

\author{Wei Chen}

\affiliation{Department of Physics, PUC-Rio, 22451-900 Rio de Janeiro, Brazil}

\date{\rm\today}

\begin{abstract}

The quantum geometry in the momentum space of semiconductors and insulators, described by the quantum metric of the valence band Bloch state, has been an intriguing issue owing to its connection to various material properties. Because the Brillouin zone is periodic, the integration of quantum metric over momentum space represents an average distance between neighboring Bloch states, of which we call the fidelity number. We show that this number can further be expressed in real space as a fidelity marker, which is a local quantity that can be calculated directly from diagonalizing the lattice Hamiltonian. A linear response theory is further introduced to generalize the fidelity number and marker to finite temperature, and moreover demonstrates that they can be measured from the global and local optical absorption power against linearly polarized light. In particular, the fidelity number spectral function in 2D systems can be easily measured from the opacity of the material. Based on the divergence of quantum metric, a nonlocal fidelity marker is further introduced and postulated as a universal indicator of any quantum phase transitions provided the crystalline momentum remains a good quantum number, and it may be interpreted as a Wannier state correlation function. The ubiquity of these concepts is demonstrated for a variety of topological insulators and topological phase transitions in different dimensions.

\end{abstract}

\maketitle

\section{Introduction}

For materials with a band gap like insulators or semiconductors, the completely filled valence bands at zero temperature define a compact manifold parameterized by the crystalline momentum ${\bf k}$ owing to the periodicity of the Brillouin zone (BZ). Recently, the notion of quantum geometry\cite{Provost80} of the valence band Bloch states $|\psi({\bf k})\rangle$ on this compact manifold emerges as an important issue that has been linked to various materials properties\cite{Ma13,Kolodrubetz13,Ma14,Yang15,Piechon16,Kolodrubetz17,Ozawa18,Palumbo18,
Palumbo18_2,Lapa19,Yu19,Chen20_Palumbo,Ma20,Salerno20,Lin21,Mitscherling20,Mitscherling22,Ahn20,Ahn22}. As in differential geometry and general relativity, the discussion of quantum geometry starts by equipping the manifold with a quantum metric $g_{\mu\nu}({\bf k})$ defined from the overlap of neighboring Bloch states $|\langle\psi({\bf k})|\psi({\bf k+\delta k})\rangle|=1-g_{\mu\nu}\delta k^{\mu}\delta k^{\nu}/2$, which can also be viewed as a fidelity susceptibility defined with respect to momentum ${\bf k}$\cite{You07,Zanardi07,Gu08,Yang08,Albuquerque10,Gu10,Carollo20}. Once the metric is defined, one can proceed to introduce various geometrical quantities like Ricci scalar, Riemann tensor, geodesics, etc., and discuss their physical interpretations. For instance, the geodesics have the physical interpretation as the trajectories in the momentum space along which the Bloch state $|\psi({\bf k})\rangle$ as a unit vector rotates the least in the Hilbert space. 


From a differential geometrical point of view, the $D$-dimensional BZ manifold is special in the sense that it is a $T^{D}$ torus, and hence it is possible to lay a single coordinate chart ${\bf k}=(k_{1},k_{2}...k_{D})$ for the entire manifold. 
It then follows that the momentum integration of the quantum metric $\int d^{D}{\bf k}\,g_{\mu\nu}({\bf k})$ over the $T^{D}$ torus represents an {\it average distance} between neighboring Bloch states $|\psi({\bf k})\rangle$ and $|\psi({\bf k+\delta k})\rangle$, thereby serving as a characteristic differential geometrical property of the manifold, which has been related to the spread of Wannier functions\cite{Souza00,Marzari97,Marzari12}. In this paper, we call this momentum integration the fidelity number, in a way completely analogous to the Chern number as a momentum integration of the Berry curvature, except the fidelity number needs not be quantized. We first use the linear response theory recently developed for the Chern number\cite{Molignini22_Chern_marker} and spin Chern number\cite{Chen23_spin_Chern} to generalize the fidelity number to finite temperature, which also recognizes it as the absorption power of the material against linearly polarized light, whose dependence on the frequency of the light is described by a spectral function. This indicates that the optical absorption measurements, which have been performed in semiconductors for decades, can actually be used to reveal the quantum geometrical properties of the material. Particularly for 2D materials, we will elaborate that this spectral function can be easily measured from the opacity of the material\cite{Nair08}.


We then proceed to convert the fidelity number into a real space object that we call the fidelity marker. This marker is introduced through drawing analogy with the theory of Chern marker\cite{Prodan10,Prodan11,Bianco11}, which maps the Chern number of 2D materials into a real space quantity through a projector formalism, and has been recently generalized to topological materials in any dimension and symmetry class\cite{Chen22_universal_marker}. In fact, this projector formalism has been applied to the spread of Wannier functions, yielding a localization marker that can be used to distinguish metals and insulators\cite{Marrazzo19}. In contrast, we will apply this projector formalism directly to the fidelity number to obtain what we call the fidelity marker, which is equivalently to a symmetrized localization marker that is experimentally measurable. Moreover, we introduce a nonlocal fidelity marker from the Fourier transform of the quantum metric, which has the physical meaning of a Wannier state correlation function, and is suggested to be a universal indicator for quantum phase transitions owing to the divergence of quantum metric, provided the crystalline momentum ${\bf k}$ remains a good quantum number.

To demonstrate the applications of the local and nonlocal fidelity markers, we turn to generic topological insulators (TIs) and topological phase transitions (TPTs) from 1D to 3D. Our survey is motivated by a recently discovered connection between the topological order and quantum metric, namely the module of the curvature functions that integrates to the topological invariant is equal to the determinant of the quantum metric\cite{Ma13,Ma14,Yang15,Piechon16,Panahiyan20_fidelity_susceptibility,Mera22}, a relation that has been called the metric-curvature correspondence\cite{vonGersdorff21_metric_curvature}. This correspondence holds for Dirac models in any dimension and symmetry class\cite{Schnyder08,Ryu10,Kitaev09,Chiu16}, as can be derived from a universal topological invariant\cite{vonGersdorff21_unification}. In addition, because the curvature function generally diverges at the gap-closing high symmetry points (HSPs) in the BZ as the material approaches a TPT\cite{Chen16,Chen16_2,Chen17,Chen19_AMS_review,Chen19_universality,MoligniniReview:2019}, the quantum metric must diverge accordingly. As a consequence, the optical absorption of the whole lattice, i.e., the fidelity number, is predicted to show some anomaly near TPTs.





The structure of the paper is organized in the following manner. In Sec.~II, we first introduce the fidelity number and marker from a projector formalism, and then use a linear response theory to generalize them to finite temperature and relate them to optical absorption power. A nonlocal fidelity marker is subsequently introduced from Fourier transform of the quantum metric. Section III gives concrete results for generic TIs in 1D, 2D, and 3D, with a clear demonstration of the singular behavior near TPTs. Finally, we summarize our results in Sec.~IV, and give an outlook on other possible applications of the fidelity marker.


\section{Mapping quantum geometry and quantum phase transitions to real space \label{sec:mapping_quantum_geometry}}


\subsection{Local fidelity marker at zero temperature \label{sec:zeroT_fidelity_marker}} 

Our aim is to elaborate that the momentum-integrated quantum metric serves as a characteristic quantum geometrical property that can be defined on lattice sites, and to construct a linear response theory to connect it to experimental measurables. Focusing on insulating materials with a band gap, we will reserve the index $n$ for valence bands, $m$ for conduction bands, $\ell$ for all the bands, and likewisely for the summations $\left\{\sum_{n},\sum_{m},\sum_{\ell}\right\}$. The $\langle{\bf r}|\ell^{\bf k}\rangle=\ell^{\bf k}({\bf r})=e^{-i{\bf k\cdot r}}\psi_{\ell}^{\bf k}({\bf r})$ is the periodic part of the Bloch state satisfying $\ell^{\bf k}({\bf r})=\ell^{\bf k}({\bf r+R})$, where ${\bf r}$ and ${\bf R}$ are Bravais lattice vectors. The Bloch state of each band $|\ell^{\bf k}\rangle$ corresponds to a Wannier state $|{\bf R}\ell\rangle$ according to
\begin{eqnarray}
|\ell^{\bf k}\rangle=\sum_{{\bf R}}e^{-i {\bf k}\cdot({\hat{\bf r}}-{\bf R})}|{\bf R}\ell\rangle,\;\;
|{\bf R} \ell\rangle=\sum_{\bf k}e^{i {\bf k}\cdot({\hat{\bf r}}-{\bf R})}|\ell^{\bf k}\rangle,\;\;\;\;\;
\label{Wannier_basis}
\end{eqnarray}
where the corresponding Wannier function $\langle {\bf r}|{\bf R} n\rangle=W_{n}({\bf r}-{\bf R})$ is highly localized around the home cell ${\bf R}$.

For a system of $N_{-}$ valence bands, the fully antisymmetric valence band Bloch state at momentum ${\bf k}$ is given by
\begin{eqnarray}
|u^{\rm val}({\bf k})\rangle=\frac{1}{\sqrt{N_{-}!}}\epsilon^{n_{1}n_{2}...n_{N-}}|n_{1}^{\bf k}\rangle|n_{2}^{\bf k}\rangle...|n_{N_{-}}^{\bf k}\rangle,\;\;\;
\label{psi_val}
\end{eqnarray}
Our interest is the quantum metric of this state defined from\cite{Provost80} 
\begin{eqnarray}
|\langle u^{\rm val}({\bf k})|u^{\rm val}({\bf k+\delta k})\rangle|=1-\frac{1}{2}g_{\mu\nu}({\bf k})\delta k^{\mu}\delta k^{\nu},
\label{uval_gmunu}
\end{eqnarray}
which amounts to the expression\cite{vonGersdorff21_metric_curvature} 
\begin{eqnarray}
&&g_{\mu\nu}({\bf k})=\frac{1}{2}\langle \partial_{\mu}u^{\rm val}|\partial_{\nu}u^{\rm val}\rangle+\frac{1}{2}\langle \partial_{\nu}u^{\rm val}|\partial_{\mu}u^{\rm val}\rangle
\nonumber \\
&&-\langle \partial_{\mu}u^{\rm val}|u^{\rm val}\rangle \langle u^{\rm val}|\partial_{\nu}u^{\rm val}\rangle
\nonumber \\
&&=\frac{1}{2}\sum_{nm}\left[\langle \partial_{\mu}n|m\rangle\langle m|\partial_{\nu}n\rangle+\langle \partial_{\nu}n|m\rangle\langle m|\partial_{\mu}n\rangle\right].
\label{gmunu_T0}
\end{eqnarray}
The key aspect of the present work is the fidelity number calculated from momentum integration of the quantum metric, that can further be expressed as overlap of Wannier states\cite{Molignini22_Chern_marker}
\begin{eqnarray}
&&{\cal G}_{\mu\nu}=\int\frac{d^{D}{\bf k}}{(2\pi)^{D}}g_{\mu\nu}({\bf k})
\nonumber \\
&&=\frac{1}{2\hbar^{2}}\int\frac{d^{D}{\bf k}}{(2\pi)^{D}}\sum_{nm}\langle \psi_{n}^{\bf k}|{\hat\mu}|\psi_{m}^{\bf k}\rangle\langle \psi_{m}^{\bf k}|{\hat\nu}|\psi_{n}^{\bf k}\rangle+(\mu\leftrightarrow\nu)
\nonumber \\
&&=\frac{\hbar^{D-2}}{2a^{D}}\int\frac{d^{D}{\bf k}}{(2\pi\hbar/a)^{D}}\int\frac{d^{D}{\bf k'}}{(2\pi\hbar/a)^{D}}
\nonumber \\
&&\times\sum_{nm}\langle \psi_{n}^{\bf k}|{\hat\mu}|\psi_{m}^{\bf k'}\rangle\langle \psi_{m}^{\bf k'}|{\hat\nu}|\psi_{n}^{\bf k}\rangle
+(\mu\leftrightarrow\nu)
\nonumber \\
&&=\sum_{nm}\sum_{\bf R}\frac{1}{2}\langle{\bf 0}n|{\hat \mu}|{\bf R}m\rangle\langle{\bf R}m|{\hat \nu}|{\bf 0}n\rangle+(\mu\leftrightarrow\nu)
\nonumber \\
&&=\sum_{nm}\sum_{\bf R}\frac{1}{2}\int d{\bf r}\int d{\bf r'}\mu_{\bf r}W_{n}^{\ast}({\bf r})W_{m}({\bf r-R})\nu_{\bf r'}
\nonumber \\
&&\times W_{m}^{\ast}({\bf r'-R})W_{n}({\bf r'})+(\mu\leftrightarrow\nu),
\label{Gmunu_definition}
\end{eqnarray}
where the third line is valid due to the fact that the matrix elements vanish if ${\bf k\neq k'}$. In deriving the above expression we have used the identity\cite{Bianco11,Marzari97,Marzari12,Gradhand12,Chen17,Chen19_AMS_review}
\begin{eqnarray}
&&i\langle m|\partial_{\mu} n\rangle=\frac{1}{\hbar}\langle \psi_{m}^{\bf k}|{\hat \mu}|\psi_{n}^{\bf k}\rangle
\nonumber \\
&&=\sum_{\bf R}\frac{e^{i{\bf k\cdot R}}}{\hbar}\langle{\bf 0}m|{\hat \mu}|{\bf R}n\rangle=\sum_{\bf R}\frac{e^{-i{\bf k\cdot R}}}{\hbar}\langle{\bf R}m|{\hat \mu}|{\bf 0}n\rangle,
\label{udu_to_psixpsi}
\end{eqnarray}
for $m\neq n$. Within the context of differential geometry, ${\cal G}_{\mu\nu}$ in Eq.~(\ref{Gmunu_definition}) measures the average distance between neighboring points on the BZ torus $T^{D}$, as can be understood from the definition in Eq.~(\ref{uval_gmunu}). Note that in systems with a zero-energy flat band, the momentum-integration of quantum metric of the flat band alone has been related to the superfluid weight\cite{Peotta15,Julku16,Rossi21,Torma21}, which may be relevant to the superconductivity in twisted bilayer graphene\cite{Xie20}. In contrast, here we consider insulators or semiconductors, and the appropriate quantum metric is that of the fully antisymmetric state described by Eq.~(\ref{psi_val}) that includes the contribution from all the valence bands.

Now suppose we have a tight-binding Hamiltonian $H=\sum_{\bf rr'\sigma\sigma '}t_{\bf rr'\sigma\sigma '}c_{\bf r\sigma}^{\dag}c_{\bf r'\sigma '}$, where ${\bf r}$ labels the lattice sites on a $D$-dimensional lattice, that has been diagonalized and the eigenstates are found via $H|E_{\ell}\rangle=E_{\ell}|E_{\ell}\rangle$. The projectors to the filled and empty lattice eigenstates can be constructed analogously from the projectors to the valence and conduction band states integrated over momentum, yielding\cite{Bianco11}
\begin{eqnarray}
&&{\hat P}=\sum_{n}\int\frac{d^{D}{\bf k}}{(2\pi\hbar/a)^{D}}|\psi_{n}^{\bf k}\rangle\langle\psi_{n}^{\bf k}|\rightarrow\sum_{n}|E_{n}\rangle\langle E_{n}|,
\nonumber \\
&&{\hat Q}=\sum_{m}\int\frac{d^{D}{\bf k'}}{(2\pi\hbar/a)^{D}}|\psi_{m}^{\bf k'}\rangle\langle\psi_{m}^{\bf k'}|\rightarrow \sum_{m}|E_{m}\rangle\langle E_{m}|.
\label{projector_PQ_k}
\end{eqnarray}
Using these projectors and following the same construction for the local Chern marker\cite{Bianco11}, the fidelity number may be written as
\begin{eqnarray}
&&{\cal G}_{\mu\nu}=\frac{\hbar^{D-2}}{a^{D}N}{\rm Tr}\left[\frac{1}{2}{\hat P}{\hat\mu}{\hat Q}{\hat \nu}+(\mu\leftrightarrow\nu)\right]
\nonumber \\
&&=\frac{\hbar^{D-2}}{a^{D}N}{\rm Tr}\left[\frac{1}{2}{\hat P}{\hat\mu}{\hat Q}{\hat \nu}{\hat P}+(\mu\leftrightarrow\nu)\right]
=\frac{1}{N}\sum_{\bf r}{\cal G}_{\mu\nu}({\bf r}), 
\label{fidelity_number_definition}
\end{eqnarray}
where $N$ is the number of unit cells. The diagonal elements in the trace yield what we call the local fidelity marker
\begin{eqnarray}
&&{\cal G}_{\mu\nu}({\bf r})=\frac{\hbar^{D-2}}{a^{D}}\sum_{\sigma}\langle{\bf r,\sigma}|\left[\frac{1}{2}{\hat P}{\hat\mu}{\hat Q}{\hat \nu}{\hat P}+(\mu\leftrightarrow\nu)\right]|{\bf r,\sigma}\rangle
\nonumber \\
&&\equiv \frac{\hbar^{D-2}}{a^{D}}\langle{\bf r}|\left[\frac{1}{2}{\hat P}{\hat\mu}{\hat Q}{\hat \nu}{\hat P}+(\mu\leftrightarrow\nu)\right]|{\bf r}\rangle
\label{fidelity_marker_definition}
\end{eqnarray}
where $\sigma$ denotes any internal degrees of freedom within a unit cell such as spin, orbit, sublattice, etc. We should call the operator ${\hat {\cal G}_{\mu\nu}}\equiv\left[{\hat P}{\hat\mu}{\hat Q}{\hat \nu}{\hat P}+{\hat P}{\hat\nu}{\hat Q}{\hat \mu}{\hat P}\right]/2$ in these equations the fidelity operator. In Eq.~(\ref{fidelity_number_definition}), we have replaced the trace by an equivalent ${\rm Tr}[{\hat P}{\hat\mu}{\hat Q}{\hat \nu}]={\rm Tr}[{\hat P}{\hat\mu}{\hat Q}{\hat \nu}{\hat P}]$, which seems to be redundant since ${\hat P}{\hat P}={\hat P}$, but it is know that this step is crucial to get the nonzero diagonal elements\cite{Bianco11}, i.e., the fidelity marker in Eq.~(\ref{fidelity_marker_definition}), as we also confirm numerically.  Interestingly, in contrast to the Chern marker that acts like the commutator of the two operators ${\hat P}{\hat\mu}{\hat Q}{\hat \nu}$ and ${\hat P}{\hat\nu}{\hat Q}{\hat \mu}$\cite{Prodan10,Prodan11,Bianco11}, the quantum metric marker is constructed from the anticommutator of the same two operators. Finally, we remark that our fidelity marker is equivalent to the previously proposed localization marker ${\cal L}_{\mu\nu}$ but symmetrized in the two indices $\mu$ and $\nu$, as elaborated in Appendix \ref{apx:comparison_localization_marker}. This symmetrization stems from the very definition of the quantum metric in Eq.~(\ref{gmunu_T0}) and makes our marker experimentally measurable, as we shall see in Sec.~\ref{sec:linear_response}.  



\subsection{Detecting quantum phase transitions by a nonlocal fidelity marker \label{sec:nonlocal_fidelity_marker}}

Recent works suggest that in addition to using the diagonal element of the fidelity operator at site ${\bf r}$ to define the local topological marker, one may use the off-diagonal matrix elements corresponding to two different lattice sites ${\bf r}$ and ${\bf r+R}$ of the same operator to define a nonlocal marker\cite{Molignini22_Chern_marker,Chen23_spin_Chern,Chen22_universal_marker}. Such a nonlocal marker is equivalently a Wannier state correlation function calculated from the Fourier transform of the curvature function\cite{Chen17,Chen19_AMS_review,Chen19_universality}. Here we demonstrate that a similar formalism gives rises to a nonlocal fidelity marker that can also be interpreted as a Wannier state correlation function, as we shall see below.

Consider the Fourier transform of the quantum metric denoted by $\tilde{g}_{\mu\nu}({\bf R})$, which can be expressed in terms of Wannier states by\cite{Molignini22_Chern_marker}
\begin{eqnarray}
&&\tilde{g}_{\mu\nu}({\bf R})=\int\frac{d^{D}{\bf k}}{(2\pi)^{D}}g_{\mu\nu}({\bf k})e^{i{\bf k\cdot R}}
\nonumber \\
&&=\frac{\hbar^{D-2}}{a^{D}}\sum_{n}\sum_{m}\int\frac{d^{D}{\bf k}}{(2\pi\hbar/a)^{D}}\int\frac{d^{D}{\bf k'}}{(2\pi\hbar/a)^{D}}
\nonumber \\
&&\times\frac{1}{2}\langle\psi_{n}^{\bf k}|{\hat\mu}|\psi_{m}^{\bf k'}\rangle\langle \psi_{m}^{\bf k'}|{\hat\nu}|\psi_{n}^{\bf k}\rangle e^{i{\bf k\cdot R}}+(\mu\leftrightarrow\nu).
\nonumber \\
\nonumber \\
&&=\sum_{nm}\sum_{\bf R_{1}}\frac{1}{2}\langle{\bf 0}n|{\hat \mu}|{\bf R_{1}}m\rangle\langle{\bf R_{1}+R}m|{\hat \nu}|{\bf 0}n\rangle+( \mu\leftrightarrow \nu)
\nonumber \\
&&=\sum_{nm}\sum_{\bf R_{1}}\frac{1}{2}\int d{\bf r}\int d{\bf r'}\mu_{\bf r}W_{n}^{\ast}({\bf r})W_{m}({\bf r-R_{1}})\nu_{\bf r'}
\nonumber \\
&&W_{m}^{\ast}({\bf r'-R_{1}-R})W_{n}({\bf r'})+(\mu\leftrightarrow\nu),
\end{eqnarray}
where we have used the fact that $|\psi_{n}^{\bf k}\rangle e^{i{\bf k\cdot R}}$ projected to $\langle{\bf r}|$ is equal to $|\psi_{n}^{\bf k}\rangle$ projected to $\langle{\bf r+R}|$ due to the cell periodicity $n({\bf r})=n({\bf r+R})$. On the other hand, if in the projector formalism we consider the $({\bf r+R,r})$-th off-diagonal element of the matrix and call it the nonlocal fidelity marker ${\cal G}_{\mu\nu}({\bf r+R,r})$
\begin{eqnarray}
&&{\cal G}_{\mu\nu}({\bf r+R,r})
\nonumber \\
&&=\frac{\hbar^{D-2}}{a^{D}}{\rm Re}\sum_{\sigma}\langle{\bf r+R,\sigma}|\left[\frac{1}{2}{\hat P}{\hat\mu}{\hat Q}{\hat \nu}{\hat P}+(\mu\leftrightarrow\nu)\right]|{\bf r,\sigma}\rangle
\nonumber \\
&&\equiv \frac{\hbar^{D-2}}{a^{D}}{\rm Re}\left\{\langle{\bf r+R}|\left[\frac{1}{2}{\hat P}{\hat\mu}{\hat Q}{\hat \nu}{\hat P}+(\mu\leftrightarrow\nu)\right]|{\bf r}\rangle\right\},
\label{nonlocal_fidelity_marker}
\end{eqnarray}
then it becomes clear that the Wannier state correlation function $\tilde{g}_{\mu\nu}({\bf R})$ is the clean, theomodynamic limit of the nonlocal fidelity marker $\lim_{N\rightarrow\infty}{\cal G}_{\mu\nu}({\bf r+R,r})=\tilde{g}_{\mu\nu}({\bf R})$.

The objective of introducing this nonlocal fidelity marker is to examine whether its spatial profile can serve as a universal indicator for any kind of quantum phase transitions. This postulate is made because the quantum metric is essentially a fidelity susceptibility defined by treating momentum ${\bf k}$ as a tuning parameter. Thus as the system approaches a quantum phase transition, the quantum metric is expected to diverge at some momentum provided that momentum remains a good quantum number\cite{You07,Zanardi07,Gu08,Yang08,Albuquerque10,Gu10,Carollo20}, and consequently the nonlocal fidelity marker as the Fourier transform of the quantum metric should become more long ranged. We shall use some concrete examples in the following sections to support this conjecture.




\subsection{Linear response theory of finite temperature fidelity number \label{sec:linear_response}}

We now generalize the fidelity number and marker to finite temperature using the linear response theory developed previously for the Chern marker and spin Chern marker\cite{Molignini22_Chern_marker,Chen23_spin_Chern}, and elaborate that it can be measured by optical absorption of the material. Remarkably, our formalism implies that the gauge-invariant part of the (symmetrized) spread of Wannier functions can be measured by optical absorption power, which may help to experimentally verify the theoretical results obtained from first principle calculations\cite{Souza00,Marzari97,Marzari12}. We start from a quantum metric spectral function derived from the response of the system under an oscillating electric field\cite{Chen22_dressed_Berry_metric} 
\begin{eqnarray}
&&g_{\mu\nu}^{d}({\bf k},\omega)=\sum_{\ell<\ell '}\left\{\frac{1}{2}\langle\partial_{\mu}\ell|\ell '\rangle\langle\ell '|\partial_{\nu}\ell\rangle+(\mu\leftrightarrow\nu)\right\}
\nonumber \\
&&\times\left[f(\varepsilon_{\ell}^{\bf k})-f(\varepsilon_{\ell '}^{\bf k})\right]\delta(\omega+\frac{\varepsilon_{\ell}^{\bf k}}{\hbar}-\frac{\varepsilon_{\ell '}^{\bf k}}{\hbar}).
\label{gmunuw_finiteT}
\end{eqnarray}
The superscript $d$ stands for "dressed" since the formalism is valid at finite temperature (and potentially including many-body interactions as well, although they will not be addressed here).

Experimentally, $g_{\mu\nu}^{d}(\omega)$ in Eq.~(\ref{gmunuw_finiteT}) is related to the optical response at momentum ${\bf k}$. This can be seen by considering the current operator in momentum space ${\hat j}_{\mu}=e\partial_{\mu}H$, where $\partial_{\mu}\equiv\partial/\partial k_{\mu}$ and $H=H({\bf k})$ is the momentum space single-particle Hamiltonian, and we will assume a $D$-dimensional cubic lattice of unit cell volume $a^{D}$. The usual linear response theory for noninteracting system gives the finite temperature longitudinal optical conductivity at momentum ${\bf k}$ at frequency of the light $\omega$ by\cite{Mahan00}
\begin{eqnarray}
&&\sigma_{\mu\mu}({\bf k},\omega)=\sum_{\ell<\ell '}\frac{\pi}{a^{D}\hbar\omega}\langle\ell|{\hat j}_{\mu}|\ell '\rangle\langle\ell '|{\hat j}_{\mu}|\ell\rangle
\nonumber \\
&&\times\left[f(\varepsilon_{\ell}^{\bf k})-f(\varepsilon_{\ell '}^{\bf k})\right]\delta(\omega+\frac{\varepsilon_{\ell}^{\bf k}}{\hbar}-\frac{\varepsilon_{\ell '}^{\bf k}}{\hbar})
\nonumber \\
&&=\frac{\pi e^{2}}{a^{D}}\hbar\omega\,g_{\mu\mu}^{d}({\bf k},\omega),
\label{sigmaw_gmumu}
\end{eqnarray}
which can be used to extract the diagonal components of the quantum metric spectral function $g_{\mu\mu}^{d}({\bf k},\omega)$. Here $|\ell\rangle\equiv |\ell^{\bf k}\rangle$ is the periodic part of the Block state, and the index $\ell$ enumerates both the valence and conduction band states, since at finite temperature both of them contribute to the conductivity. The $f(\varepsilon_{\ell}^{\bf k})$ is the Fermi distribution of the eigenenergy $\varepsilon_{\ell}^{\bf k}$. Moreover, because the optical conductivity corresponds to the optical absorption process, the $\delta(\omega+\varepsilon_{\ell}^{\bf k}/\hbar-\varepsilon_{\ell '}^{\bf k}/\hbar)$ with $\omega>0$ ensures that $\varepsilon_{\ell}^{\bf k}<\varepsilon_{\ell '}^{\bf k}$ as denoted by the summation $\sum_{\ell <\ell '}$. Many previous works\cite{Souza00,Resta02,Resta11,Chen22_dressed_Berry_metric,Kashihara22} have already pointed out that the frequency integration of this spectral function gives the quantum metric $g_{\mu\mu}^{d}({\bf k})=\int_{0}^{\omega}d\omega\,g_{\mu\mu}^{d}({\bf k},\omega)$ that in the zero temperature limit $\lim_{T\rightarrow 0}g_{\mu\nu}^{d}({\bf k})=g_{\mu\nu}({\bf k})$ recovers the expression in Eq.~(\ref{gmunu_T0}). In contrast, we will focus on the spetral function $g_{\mu\mu}^{d}({\bf k},\omega)$ itself for the purpose that will become clear below.

The conductivity of the whole sample, which is also what is measured in real space, is given by the momentum integration of the above quantity
\begin{eqnarray}
&&\sigma_{\mu\mu}(\omega)=\int\frac{d^{D}{\bf k}}{(2\pi\hbar/a)^{D}}\,\sigma_{\mu\mu}({\bf k},\omega)
\nonumber \\
&&=\frac{\pi e^{2}}{\hbar^{D-1}}\,\omega\int\frac{d^{D}{\bf k}}{(2\pi)^{D}}\,g_{\mu\mu}^{d}({\bf k},\omega)
\equiv\frac{\pi e^{2}}{\hbar^{D-1}}\,\omega\,{\cal G}_{\mu\mu}^{d}(\omega),
\label{fidelity_number_spec_fn}
\end{eqnarray}
which defines what we call the fidelity number spectral function ${\cal G}_{\mu\mu}^{d}(\omega)$. On the other hand, in a $D$-dimension material, if we denote the polarized oscillating field by $E_{\mu}(\omega,t)=E_{0}\cos\omega t$ and the current that it induces by $j_{\mu}(\omega,t)=\sigma_{\mu\mu}(\omega)E_{0}\cos\omega t$, where $E_{0}$ is the strength of the field, then the optical absorption power per unit cell at frequency $\omega$ is 
\begin{eqnarray}
&&W_{a}(\omega)=\langle j_{\mu}(\omega,t)E_{\mu}(\omega,t)\rangle_{t}=\frac{1}{2}\sigma_{\mu\mu}(\omega)E_{0}^{2}
\nonumber \\
&&=\frac{\pi e^{2}}{2\hbar^{D-1}}\,E_{0}^{2}\omega\,{\cal G}_{\mu\mu}^{d}(\omega),
\label{absorption_power_global}
\end{eqnarray}
which can be used to extract the spectral function ${\cal G}_{\mu\mu}^{d}(\omega)$, where $\langle...\rangle_{t}$ denotes the time average. Particularly in 2D systems, the incident power of the light per unit area is $W_{i}=c\varepsilon_{0}E_{0}^{2}/2$, and hence the opacity of the 2D system as the incident light has frequency $\omega$ and polarization $\mu$ is\cite{Nair08} 
\begin{eqnarray}
{\cal O}(\omega)=\frac{W_{a}(\omega)}{W_{i}}=4\pi^{2}\alpha\omega\,{\cal G}_{\mu\mu}^{d}(\omega)|_{2D},
\end{eqnarray}
where $\alpha=e^{2}/4\pi\varepsilon_{0}\hbar c$ is the fine structure constant. Thus the fidelity number spectral function can be simply extracted experimentally from the opacity by
\begin{eqnarray}
{\cal G}_{\mu\mu}^{d}(\omega)|_{2D}=\frac{1}{4\pi\omega}\left[\frac{{\cal O}(\omega)}{\pi\alpha}\right].
\label{fidelity_number_opacity}
\end{eqnarray}
In other words, it is simply the opacity at $\omega$ and polarization $\mu$ measured in units of $\pi\alpha\approx 2.3\%$ and then divided by $4\pi\omega$. We expect that this simple protocol should be broadly applicable to 2D systems, as has been elaborated recently for 2D Dirac materials like graphene and silicene\cite{deSousa23_graphene_opacity}.

To extract the off-diagonal components of the fidelity number spectral function, such as ${\cal G}_{xy}^{d}(\omega)$, one may consider linearly polarized light with a phase different\cite{Ozawa18} ${\bf E}_{\pm}=E_{0}({\hat{\bf x}}\pm{\hat{\bf y}})$, and the current operators ${\hat j}_{\pm}={\hat j}_{x}\pm {\hat j}_{y}$ and conductivities $\sigma_{\pm}({\bf k},\omega)$ defined accordingly. Using the off-diagonal component $g_{xy}^{d}({\bf k},\omega)$ in Eq.~(\ref{gmunuw_finiteT}), the same analysis leads to the conclusion that the difference of optical conductivity between the two polarizations gives 
\begin{eqnarray}
&&\sigma_{+}({\bf k},\omega)-\sigma_{-}({\bf k},\omega)=4\frac{\pi e^{2}}{a^{D}}\hbar\omega\,g_{xy}^{d}({\bf k},\omega),
\end{eqnarray}
and after a momentum integration, the difference in absorption power gives
\begin{eqnarray}
&&\Delta W_{a}(\omega)=\langle j_{+}(\omega,t)E_{+}(\omega,t)-j_{-}(\omega,t)E_{-}(\omega,t)\rangle
\nonumber \\
&&=2\frac{\pi e^{2}}{\hbar^{D-1}}\,E_{0}^{2}\omega\,{\cal G}_{xy}^{d}(\omega),
\label{DeltaWa_Gxy}
\end{eqnarray}
which may be used to extract the off-diagonal component ${\cal G}_{xy}^{d}(\omega)$. For 2D systems, the incident power of the two polarizations we considered is $W_{i}^{\pm}=c\varepsilon_{0}E_{0}^{2}|{\hat{\bf x}}\pm{\hat{\bf y}}|^{2}/2=c\varepsilon_{0}E_{0}^{2}$, and hence
\begin{eqnarray}
{\cal G}_{xy}^{d}(\omega)|_{2D}=\frac{1}{8\pi\omega}\left[\frac{{\cal O}_{+}(\omega)-{\cal O}_{-}(\omega)}{\pi\alpha}\right].
\label{opacity_Gxy}
\end{eqnarray}
indicating that one may use the difference in opacity under the two polarizations to extract ${\cal G}_{xy}^{d}(\omega)$.

\subsection{Finite temperature fidelity marker \label{sec:finiteT_fidelity_marker}}

After the spectral function as a function of frequency $\omega$ is obtained, the dressed fidelity number can further be obtained via a frequency integration, which can be expressed in terms of eigenstates $|E_{\ell}\rangle$ of lattice models using the procedure outlined in Sec.~\ref{sec:zeroT_fidelity_marker}, rendering 
\begin{eqnarray}
&&{\cal G}_{\mu\nu}^{d}=\int_{0}^{\infty}d\omega\,{\cal G}_{\mu\nu}^{d}(\omega)
\nonumber \\
&&=\int\frac{d^{D}{\bf k}}{(2\pi)^{D}}\sum_{\ell <\ell '}\left[\frac{1}{2}\langle\partial_{\mu}\ell|\ell '\rangle\langle\ell '|\partial_{\nu}\ell\rangle+(\mu\leftrightarrow \nu)\right]
\nonumber \\
&&\times\left[f(\varepsilon_{\ell}^{\bf k})-f(\varepsilon_{\ell '}^{\bf k})\right]
\nonumber \\
&&=\frac{1}{\hbar^{2}}\int\frac{d^{D}{\bf k}}{(2\pi)^{D}}\sum_{\ell <\ell '}\left[\frac{1}{2}\langle\psi_{\ell}^{\bf k}|{\hat\mu}|\psi_{\ell '}^{\bf k}\rangle\langle\psi_{\ell '}^{\bf k}|{\hat\nu}|\psi_{\ell}^{\bf k}\rangle+(\mu\leftrightarrow \nu)\right]
\nonumber \\
&&\times\left[f(\varepsilon_{\ell}^{\bf k})-f(\varepsilon_{\ell '}^{\bf k})\right]
\nonumber \\
&&=\frac{\hbar^{D-2}}{a^{D}}\int\frac{d^{D}{\bf k}}{(2\pi\hbar/a)^{D}}\int\frac{d^{D}{\bf k '}}{(2\pi\hbar/a)^{D}}\sum_{\ell <\ell '}
\nonumber \\
&&\times\left[\frac{1}{2}\langle\psi_{\ell}^{\bf k}|{\hat\mu}|\psi_{\ell '}^{\bf k'}\rangle\langle\psi_{\ell '}^{\bf k'}|{\hat\nu}|\psi_{\ell}^{\bf k}\rangle+(\mu\leftrightarrow \nu)\right]
\nonumber \\
&&\times\left[f(\varepsilon_{\ell}^{\bf k})-f(\varepsilon_{\ell '}^{\bf k'})\right]
\nonumber \\
&&=\frac{\hbar^{D-2}}{Na^{D}}\sum_{\ell <\ell '}\left[\frac{1}{2}\langle E_{\ell}|{\hat\mu}|E_{\ell '}\rangle\langle E_{\ell '}|{\hat\nu}|E_{\ell}\rangle+(\mu\leftrightarrow \nu)\right]
\nonumber \\
&&\times\left[f(E_{\ell})-f(E_{\ell '})\right]
\nonumber \\
&&=\frac{\hbar^{D-2}}{Na^{D}}\sum_{\ell <\ell '}{\rm Tr}\left[\frac{1}{2}{\hat \mu}S_{\ell '}{\hat \nu}S_{\ell}+(\mu\leftrightarrow \nu)\right]
\left[f(E_{\ell})-f(E_{\ell '})\right]
\nonumber \\
\label{finiteT_fidelity_number}
\end{eqnarray}
Here $|\psi_{\ell}^{\bf k}\rangle$ is the full Bloch state and $|\ell\rangle\equiv |\ell^{\bf k}\rangle$ is the periodic part of the Block state satisfying $\langle{\bf r}|\psi_{\ell}^{\bf k}\rangle=e^{i{\bf k\cdot r}}\langle{\bf r}|\ell\rangle$. The fidelity number is then expressed to the real space version using the projectors 
\begin{eqnarray}
&&\int\frac{d^{D}{\bf k}}{(2\pi\hbar/a)^{D}}|\psi_{\ell}^{\bf k}\rangle\langle\psi_{\ell}^{\bf k}|\rightarrow\sum_{\ell}|E_{\ell}\rangle\langle E_{\ell}|=\sum_{\ell}S_{\ell},
\nonumber \\
&&\int\frac{d^{D}{\bf k}}{(2\pi\hbar/a)^{D}}|\psi_{\ell}^{\bf k}\rangle\langle\psi_{\ell}^{\bf k}|f(\varepsilon_{\ell}^{\bf k})\rightarrow\sum_{\ell}|E_{\ell}\rangle\langle E_{\ell}|f(E_{\ell})
\nonumber \\
&&=\sum_{\ell}S_{\ell}f(E_{\ell}),
\end{eqnarray}
where $|E_{\ell}\rangle$ is a lattice eigenstate obtained from diagonalizing the lattice Hamiltonian $H|E_{\ell}\rangle=E_{\ell}|E_{\ell}\rangle$, and we denote its projector by $S_{\ell}=|E_{\ell}\rangle\langle E_{\ell}|$. At zero temperature $f(E_{\ell})=\theta(-E_{\ell})$, it is evident that the dressed fidelity number recovers the zero temperature one $\lim_{T\rightarrow 0}{\cal G}_{\mu\nu}^{d}={\cal G}_{\mu\nu}$, as can be easily seen by comparing Eqs.~(\ref{finiteT_fidelity_number}) and (\ref{fidelity_number_definition}).

Our goal now is to construct a finite temperature fidelity marker that spatially sums to the fidelity number ${\cal G}_{\mu\mu}^{d}=\sum_{\bf r}{\cal G}_{\mu\mu}^{d}({\bf r})/N$ in Eq.~(\ref{finiteT_fidelity_number}). To achieve this, the issue now is that according to what discussed after Eq.~(\ref{fidelity_marker_definition}), one has to add an extra projector ${\hat P}$ to the last line of Eq.~(\ref{finiteT_fidelity_number}) in order to get the correct fidelity marker. However, at finite temperature, thermal broadening renders the notion of filled and empty states rather ambiguous, let alone their projectors ${\hat P}$ and ${\hat Q}$ in Eq.~(\ref{projector_PQ_k}). To incorporate this extra projector at finite temperature, we propose to first calculate the matrix 
\begin{eqnarray}
&&{\cal M}_{\mu}=\sum_{\ell<\ell '}S_{\ell}{\hat \mu}S_{\ell '}\sqrt{f_{\ell\ell '}}.
\end{eqnarray}
where $f_{\ell\ell '}\equiv f(E_{\ell})-f(E_{\ell '})$, and then define the fidelity marker by 
\begin{eqnarray}
&&{\cal G}_{\mu\nu}^{d}({\bf r})=\frac{\hbar^{D-2}}{2a^{D}}{\rm Re}\left\{\langle{\bf r}|\left[{\cal M}_{\mu}{\cal M}_{\nu}^{\dag}+{\cal M}_{\nu}{\cal M}_{\mu}^{\dag}\right]|{\bf r}\rangle\right\}.
\nonumber \\
\label{finiteT_fidelity_marker_XY}
\end{eqnarray}
which spatially sums to the fidelity number ${\cal G}_{\mu\nu}^{d}=\sum_{\bf r}{\cal G}_{\mu\nu}^{d}({\bf r})/N$ because $\sum_{\bf r}|{\bf r}\rangle\langle{\bf r}|=I$ and $S_{\overline{\ell}'}S_{\ell '}=\delta_{\overline{\ell}'\ell '}S_{\ell '}$, and moreover recovers the zero temperature one $\lim_{T\rightarrow 0}{\cal G}_{\mu\nu}^{d}({\bf r})={\cal G}_{\mu\nu}({\bf r})$ in Eq.~(\ref{fidelity_marker_definition}), and therefore serves our purpose. Likewisely, the nonlocal fidelity marker can also be generalized to finite temperature as the off-diagonal element of this operator 
\begin{eqnarray}
&&{\cal G}_{\mu\nu}^{d}({\bf r+R,r})
\nonumber \\
&&=\frac{\hbar^{D-2}}{2a^{D}}{\rm Re}\left\{\langle{\bf r+R}|\left[{\cal M}_{\mu}{\cal M}_{\nu}^{\dag}+{\cal M}_{\nu}{\cal M}_{\mu}^{\dag}\right]|{\bf r}\rangle\right\},
\end{eqnarray}
which represents the Fourier transform of the finite temperature quantum metric $g_{\mu\nu}^{d}=\int_{0}^{\infty}d\omega\,g_{\mu\nu}^{d}(\omega)$ (defined in Eq.~(\ref{sigmaw_gmumu})) that may be used to examine the finite temperature behavior near quantum phase transitions.



It is also convenient to introduce a fidelity marker spectral function from the frequency-dependent matrix 
\begin{eqnarray}
&&{\cal M}_{\mu}(\omega)=\sum_{\ell<\ell '}S_{\ell}{\hat \mu}S_{\ell '}\sqrt{f_{\ell\ell '}\delta(\omega+E_{\ell}-E_{\ell '})},
\end{eqnarray}
where the $\delta$-function is simulated by a Lorentzian $\delta(x)=\eta/\pi(x^{2}+\eta^{2})$ in practice, so the interpretation of its square root is straightforward. The fidelity marker spectral function is then defined by
\begin{eqnarray}
&&{\cal G}_{\mu\nu}^{d}({\bf r},\omega)
\nonumber \\
&&=\frac{\hbar^{D-2}}{2a^{D}}{\rm Re}\left\{\langle{\bf r}|\left[{\cal M}_{\mu}(\omega){\cal M}_{\nu}^{\dag}(\omega)+{\cal M}_{\nu}(\omega){\cal M}_{\mu}^{\dag}(\omega)\right]|{\bf r}\rangle\right\}.
\nonumber \\
\label{fidelity_marker_spec_fn}
\end{eqnarray}
which spatially sums to the fidelity number spectral function ${\cal G}_{\mu\nu}^{d}(\omega)=\sum_{\bf r}{\cal G}_{\mu\nu}^{d}({\bf r},\omega)/N$ in Eq.~(\ref{fidelity_number_spec_fn}). Analogous to Eq.~(\ref{absorption_power_global}), this spectral function corresponds to the local optical absorption power of the unit cell at ${\bf r}$
\begin{eqnarray}
W_{a}({\bf r},\omega)=\frac{\pi e^{2}}{2\hbar^{D-1}}\,E_{0}^{2}\omega\,{\cal G}_{\mu\mu}^{d}({\bf r},\omega),
\label{absorption_power_local}
\end{eqnarray}
and likewisely the off-diagonal component ${\cal G}_{xy}^{d}({\bf r},\omega)$ corresponds to the absorption power difference in Eq.~(\ref{DeltaWa_Gxy}) defined locally at ${\bf r}$. Given this connection to the absorption power, we anticipate that thermal probes that can detect the heating caused by the light down to atomic scale, such as scanning thermal microscopy\cite{Majumdar99,Kittel05,Gomes15,Zhang20}, may be able to detect the local absorption power $W_{a}({\bf r},\omega)$ and subsequently extract ${\cal G}_{\mu\mu}^{d}({\bf r},\omega)$.


\section{Applications to topological insulators and topological phase transitions}

As a concrete example, we investigate the generic $D$-dimensional TIs in any symmetry class described by $N\times N$ Dirac models $H=\sum_{i=0}^{D}d_{i}\Gamma_{i}$, where $\Gamma_{i}$ are the Dirac matrices, and $d_{i}$ characterizes the momentum dependence of the Hamiltonian\cite{Schnyder08,Ryu10,Chiu16}. In this case, it has been proved that the fully antisymmetric valence band Bloch state of the TI at momentum ${\bf k}$, described by Eq.~(\ref{psi_val}), gives the quantum metric\cite{vonGersdorff21_metric_curvature} 
\begin{eqnarray}
g_{\mu\nu}({\bf k})=\frac{N}{8d^{2}}\left\{\sum_{i=0}^{D}\partial_{\mu}d_{i}\partial_{\nu}d_{i}
-\partial_{\mu}d\partial_{\nu}d\right\}.
\label{quantum_metric_general_formula}
\end{eqnarray}
Denoting $d=\sqrt{\sum_{i}d_{i}^{2}}$, the model has $N/2$-fold degenerate valence band states of energy $\varepsilon_{n}=-d$ and conduction band states of energy $\varepsilon_{m}=d$. The diagonal elements of fidelity number of this model have been considered in a previous work under the name of Marzari-Vanderbilt cumulant\cite{Matsuura10}. In contrast, we will emphasize the frequency- and temperature-dependence of the spectral function to draw relevance to the optical absorption power. Because the model is homogeneous, the fidelity number spectral function in Eq.~(\ref{fidelity_number_spec_fn}) and fidelity marker spectral function in Eq.~(\ref{fidelity_marker_spec_fn}) are the same
\begin{eqnarray}
&&{\cal G}_{\mu\nu}^{d}(\omega)
=\int\frac{d^{D}{\bf k}}{(2\pi)^{D}}\,g_{\mu\nu}\left[f(\varepsilon_{n})-f(\varepsilon_{m})\right]
\nonumber \\
&&\times\delta(\omega+\varepsilon_{n}/\hbar-\varepsilon_{m}/\hbar).
\label{Dirac_Gmunu_specfn}
\end{eqnarray}
We will first address the analytical results of linear Dirac models, and then investigate more realistic lattice models by numerical calculations.

\begin{figure*}[ht]
\begin{center}
\includegraphics[clip=true,width=1.99\columnwidth]{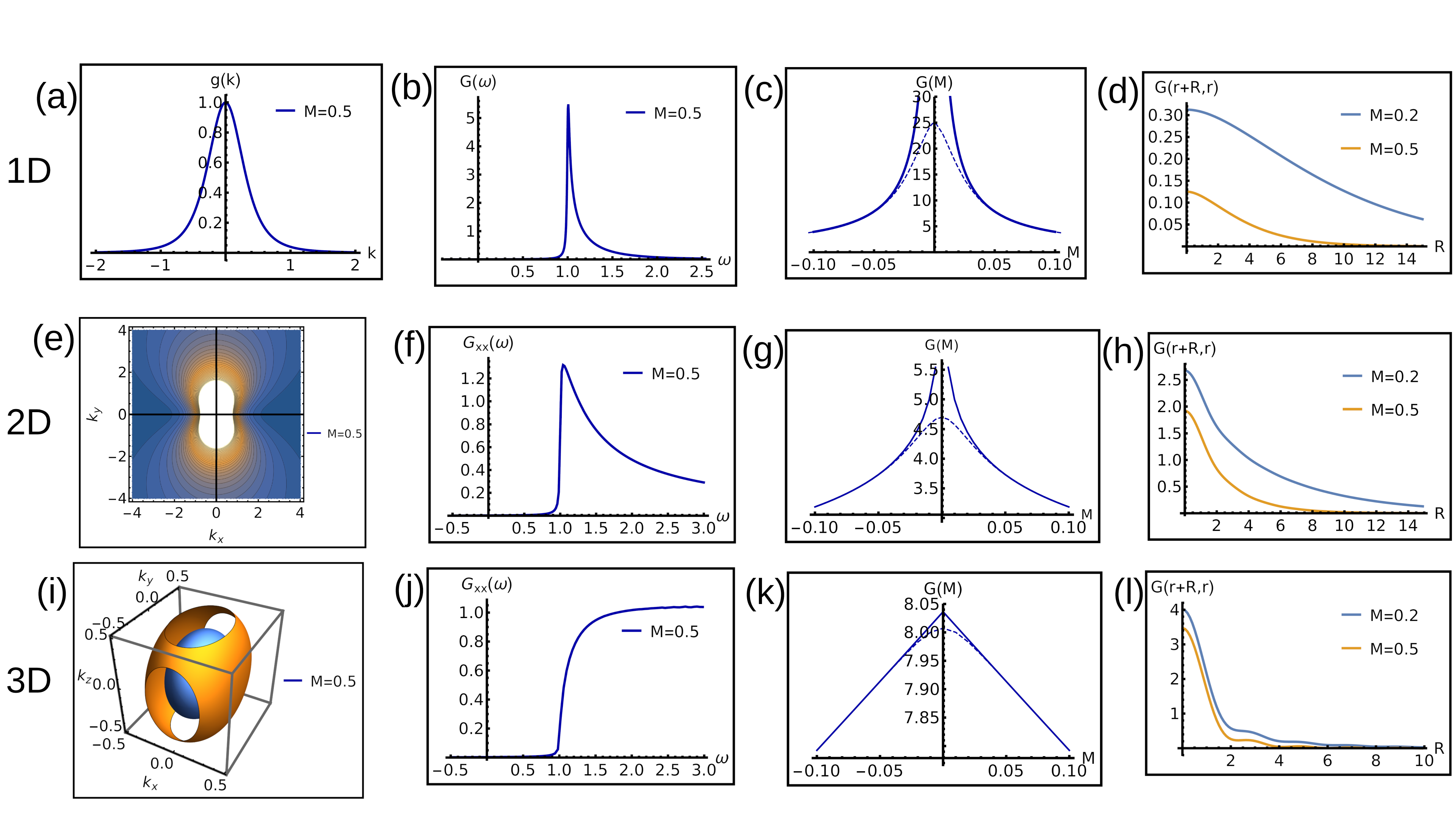}
\caption{Analytical results for homogeneous linear Dirac models of TIs in 1D (top row), 2D (middle row), and 3D (bottom row), where in (a), (e), and (i) we show the momentum profile of the quantum metric (in (i) we plot the momentum surfaces at two constant values), in (b), (f), (j) the fidelity number/marker spectral function, in (c), (g), (k) the fidelity number/marker near TPTs at zero (solid lines) and a finite (dotted lines) temperatures, and in (d), (h), (l) the spatial profile of the nonlocal fidelity marker for a parameter close to (blue) and far away from (yellow) the critical point. } 
\label{fig:continuous_model_results}
\end{center}
\end{figure*}

\subsection{Continuous models of topological insulators \label{sec:continuous_models}}

Analytical expression of ${\cal G}_{\mu\nu}^{d}(\omega)$ can be given for linear Dirac models parametrized by $d_{0}=M$ and $d_{i\neq 0}=vk_{i}$, where $v$ is the Fermi velocity. These models well describe the critical behavior of TIs and TSCs near TPTs. Using Eq.~(\ref{quantum_metric_general_formula}), the quantum metric of these models is\cite{Matsuura10}
\begin{eqnarray}
g_{\mu\nu}=\frac{N}{8d^{2}}\left(v^{2}\delta_{\mu\nu}-\frac{v^{4}k_{\mu}k_{\nu}}{d^{2}}\right).
\label{quantum_metric_linear_Dirac}
\end{eqnarray}
We will focus on the diagonal component $g_{\mu\mu}$, whose momentum profile is shown in Fig.~\ref{fig:continuous_model_results} (a), (e), and (i) for 1D, 2D, and 3D, respectively. In any dimension, the $g_{\mu\mu}$ near the gap-closing point ${\bf k=0}$ has a Lorentzian shape
\begin{eqnarray}
\lim_{k\rightarrow 0}g_{\mu\mu}=\frac{N}{8}\frac{v^{2}/M^{2}}{1+2\frac{v^{2}}{M^{2}}k^{2}},
\label{gmumu_Lorentzian}
\end{eqnarray}
whose height diverges like $g_{\mu\mu}({\bf 0})\sim 1/M^{2}$ and width vanishes like $1/\xi\sim |M|$ as the system approaches the TPT $M\rightarrow 0$, manifesting the divergence behavior anticipated at the end of Sec.~\ref{sec:nonlocal_fidelity_marker}, with a critical exponent $\nu=1$. The integration over momentum gives the diagonal components of the fidelity marker spectral function 
\begin{eqnarray}
&&{\cal G}_{\mu\mu}^{d}(\omega)=\frac{N\Omega_{D}}{2^{D+3}\pi^{D}v^{D-2}}
\left\{\frac{1}{\omega}\left[\frac{\hbar^{2}\omega^{2}}{4}-M^{2}\right]^{\frac{D}{2}-1}\right.
\nonumber \\
&&\left.-\frac{4}{D\hbar^{2}\omega^{3}}\left[\frac{\hbar^{2}\omega^{2}}{4}-M^{2}\right]^{\frac{D}{2}}\right\}
\nonumber \\
&&\times\left[f\left(-\frac{\hbar\omega}{2}\right)-f\left(\frac{\hbar\omega}{2}\right)\right]_{\omega\geq 2|M|/\hbar}
\end{eqnarray}
where $\Omega_{D}=2\pi^{\frac{D}{2}}/\Gamma(\frac{D}{2})$ for $D>1$, and $\Omega_{1}=1$.
Explicitly from 1D to 3D, it has the expressions
\begin{eqnarray}
{\cal G}_{\mu\mu}^{d}(\omega)=\left\{\begin{array}{l}
\frac{NM^{2}v}{4\pi\hbar^{2}\omega^{3}\sqrt{\hbar^{2}\omega^{2}/4-M^{2}}}
Z(\omega)|_{\omega\geq 2|M|/\hbar},\;\;\;\;\;{\rm in}\;1D \\
\left[\frac{N}{32\pi\omega}+\frac{N}{8\pi\hbar^{2}}\frac{M^{2}}{\omega^{3}}\right]Z(\omega)|_{\omega\geq 2|M|/\hbar},\;\;\;\;\;\;{\rm in}\;2D \\
\left\{\frac{NM^{2}\sqrt{\hbar^{2}\omega^{2}/4-M^{2}}}{4\pi^{2}v\hbar^{2}\omega^{3}}
+\frac{N(\hbar^{2}\omega^{2}/4-M^{2})^{3/2}}{6\pi^{2} v\hbar^{2}\omega^{3}}\right\} \\
\times Z(\omega)|_{\omega\geq 2|M|/\hbar},\;\;\;\;\;\;\;\;\;\;\;\;\;\;\;\;\;\;\;\;\;\;\;\;\;\;\;\;{\rm in}\;3D
\end{array}
\right.
\nonumber \\
\end{eqnarray}
where $Z(\omega)\equiv\left[f\left(-\frac{\hbar\omega}{2}\right)-f\left(\frac{\hbar\omega}{2}\right)\right]$, and the off-diagonal components $\mu\neq\nu$ vanish after the angular integration. The frequency dependence of ${\cal G}_{\mu\mu}^{d}(\omega)$ at zero temperature is shown in Fig.~\ref{fig:continuous_model_results} (b), (f), (j) for the 1D, 2D and 3D cases. The spectral function is practically zero at frequency lower than the band gap, as expected. Moreover, the frequency dependence above the band gap highly depends on the dimension, which gives a concrete prediction to the low frequency behavior of the optical absorption in Dirac models.

Particularly at zero temperature $f\left(-\frac{\hbar\omega}{2}\right)-f\left(\frac{\hbar\omega}{2}\right)=1$, after performing a frequency integration ${\cal G}_{\mu\mu}^{d}=\int_{2|M|/\hbar}^{\omega_{cut}}d\Omega,{\cal G}_{\mu\mu}^{d}(\omega)$ of the spectral function from the band gap $2|M|/\hbar$ to some cut-off frequency $\omega_{cut}$ that represents the band width, we obtain
\begin{eqnarray}
{\cal G}_{\mu\mu}^{d}\sim\left\{\begin{array}{l}
1/|M|\;\;\;{\rm in}\;1D, \\
\log|M|\;\;\;{\rm in}\;2D, \\
|M|+{\rm constant}\;\;\;{\rm in}\;3D,
\end{array}\right.
\label{Gmumu_analytical_critical_behavior}
\end{eqnarray}
in agreement with the previous calculation of Marzari-Vanderbilt cumulant\cite{Matsuura10}. In short, the fidelity marker diverges in 1D and 2D as the system approaches the critical point $M\rightarrow 0$ of a TPT, while saturates to a constant in 3D, as indicated in Fig.~\ref{fig:continuous_model_results} (c), (g), and (k), which also show that the effect of finite temperature is to smear out the anomaly at the critical point. Finally, in Fig.~\ref{fig:continuous_model_results} (d), (h), and (l) we show the nonlocal fidelity marker obtained from numerically performing a Fourier transform on the quantum metric. In any spatial dimension, one sees that the decay length of the nonlocal marker increases as the system approaches the critical point $M_{c}=0$, consistent with the diverging quantum metric discussed after Eq.~(\ref{gmumu_Lorentzian}). Note that for the Lorentzian shape in Eq.~(\ref{gmumu_Lorentzian}) with correlation length $\xi=\sqrt{2}|v/M|$, its Fourier transform in 1D at large $R$ gives a simple exponential decay $\tilde{g}_{\mu\mu}({\bf R})\propto e^{-R/\xi}$, but in 3D the Fourier transform $\tilde{g}_{\mu\mu}({\bf R})\propto e^{-R/\xi}/R$ has an extra factor of $1/R$, making the nonlocal marker in Fig.~\ref{fig:continuous_model_results} (l) look more short-ranged in comparison with the exponential decay in Fig.~\ref{fig:continuous_model_results} (d).

\subsection{Lattice models of topological insulators}

In this section, we use prototype lattice models of TIs in 1D, 2D, and 3D to demonstrate the real space profiles of the local and nonlocal fidelity markers. We will adopt the recipe in Sec.~\ref{sec:mapping_quantum_geometry} to directly calculate the fidelity marker on every lattice site by diagonalizing the lattice Hamiltonian in the periodic boundary condition (PBC). Note that although we only calculate the marker for some specific models, one should keep in mind that the critical behavior of the marker for all the TIs in the same dimension should be the same according to Eq.~(\ref{Gmumu_analytical_critical_behavior}).

\begin{figure}[ht]
\begin{center}
\includegraphics[clip=true,width=0.99\columnwidth]{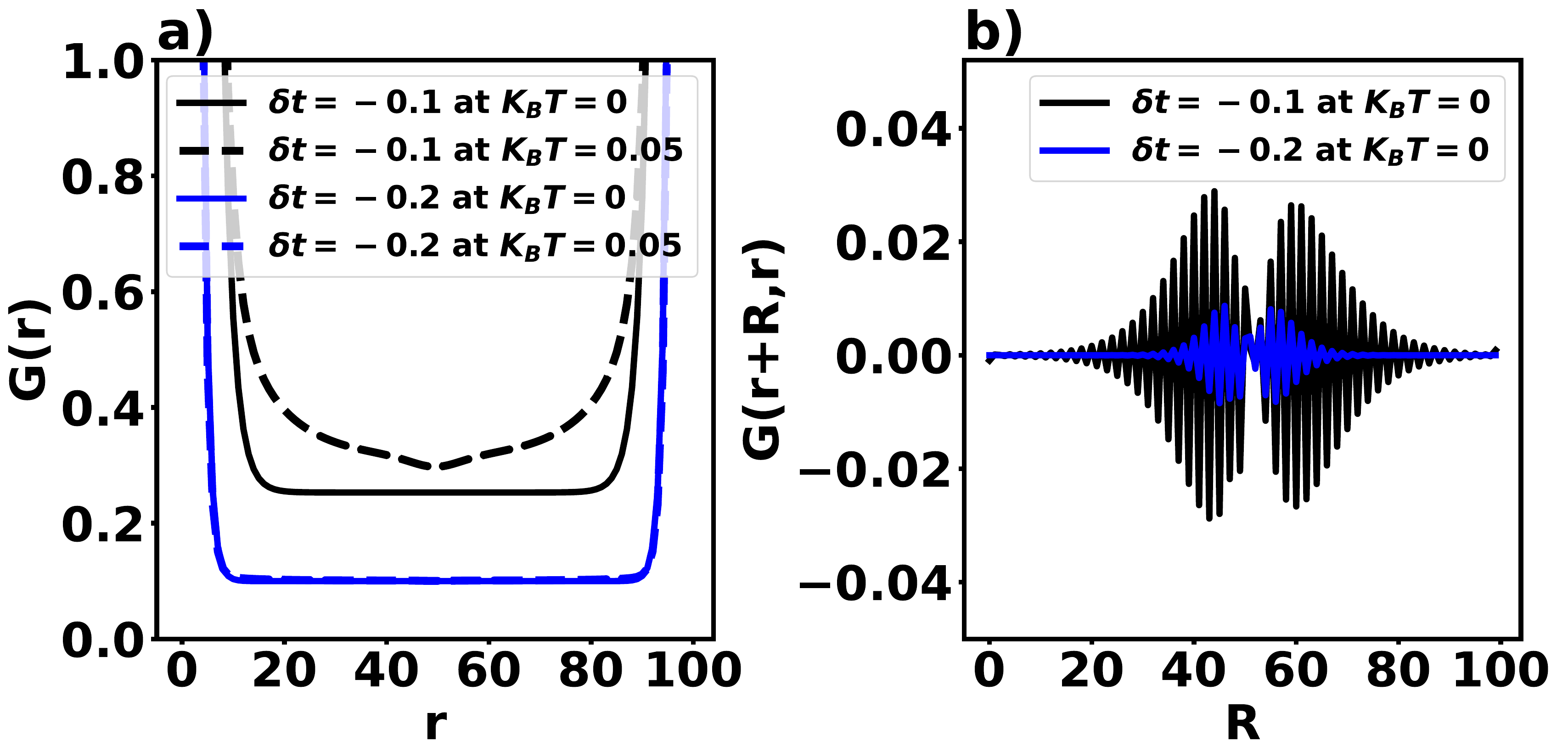}
\caption{Numerical results of 1D SSH model, where we investigate (a) the fidelity marker ${\cal G}_{xx}(r)$ at far from $\delta t=-0.2$ and close to $\delta t=-0.1$ (in units of the uniform hopping $t=1$) the critical point, and at zero and finite temperatures, and (b) the nonlocal fidelity marker at far from and close to the critical point.} 
\label{fig:1DSSH_results}
\end{center}
\end{figure}

\subsubsection{1D Su-Schrieffer-Heeger model}

A prototype example in 1D is the spinless Su-Shrieffer-Heeger (SSH) model\cite{Su79} described by the Hamiltonian
\begin{eqnarray}
H&=&\sum_{i}(t+\delta t)c_{Ai}^{\dag}c_{Bi}+(t-\delta t)c_{Ai+1}^{\dag}c_{Bi}+h.c.
\label{SSH_Hamiltonian}
\end{eqnarray}
where $c_{Ai}$ and $c_{Bi}$ are fermion annihilation operators on the $A$ and $B$ sublattices in the unit cell at $i$, and $\delta t$ is the difference between the hopping on the even and odd bonds that controls the topological order. The numerical result for the fidelity marker is shown in Fig.~\ref{fig:1DSSH_results} (a) for a lattice of 100 unit cells. Deep inside the bulk, the marker remains a constant value that agrees with the momentum-integration of quantum metric in Eq.~(\ref{Gmunu_definition}), and the value increases as the system approaches the critical point $\delta t\rightarrow 0$, in agreement with Eq.~(\ref{Gmumu_analytical_critical_behavior}) and that presented in Fig.~\ref{fig:continuous_model_results} (c) for the low energy Dirac model. Near the boundary sites, however, the marker starts to deviate from the constant value even if PBC is imposed, a problem known for this type of projector formalism since the position operator ${\hat x}$ in Eq.~(\ref{fidelity_marker_definition}) does not respect the translational invariance\cite{Bianco11}. In Fig.~\ref{fig:1DSSH_results} (b), we show that the nonlocal marker decays with distance $R$, with a decay length that grows as the system approaches the critical point $\delta t\rightarrow 0$, in agreement with that discussed at the end of Sec.~\ref{sec:nonlocal_fidelity_marker}. The maximal value of the nonlocal marker at $R=0$ recovers the local marker (setting ${\bf R=0}$ in Eq.~(\ref{nonlocal_fidelity_marker}) recovers Eq.~(\ref{fidelity_marker_definition})), i.e., the maxima of Fig.~\ref{fig:1DSSH_results} (b) are equal to the flat values in Fig.~\ref{fig:1DSSH_results} (a). In addition, the nonlocal marker also oscillates with a wave length of 2 lattice constants, owing to the fact that the nonlocal marker is the Fourier transform of the quantum metric that peaks at $k=\pi$ in this model.


\begin{figure}[ht]
\begin{center}
\includegraphics[clip=true,width=0.99\columnwidth]{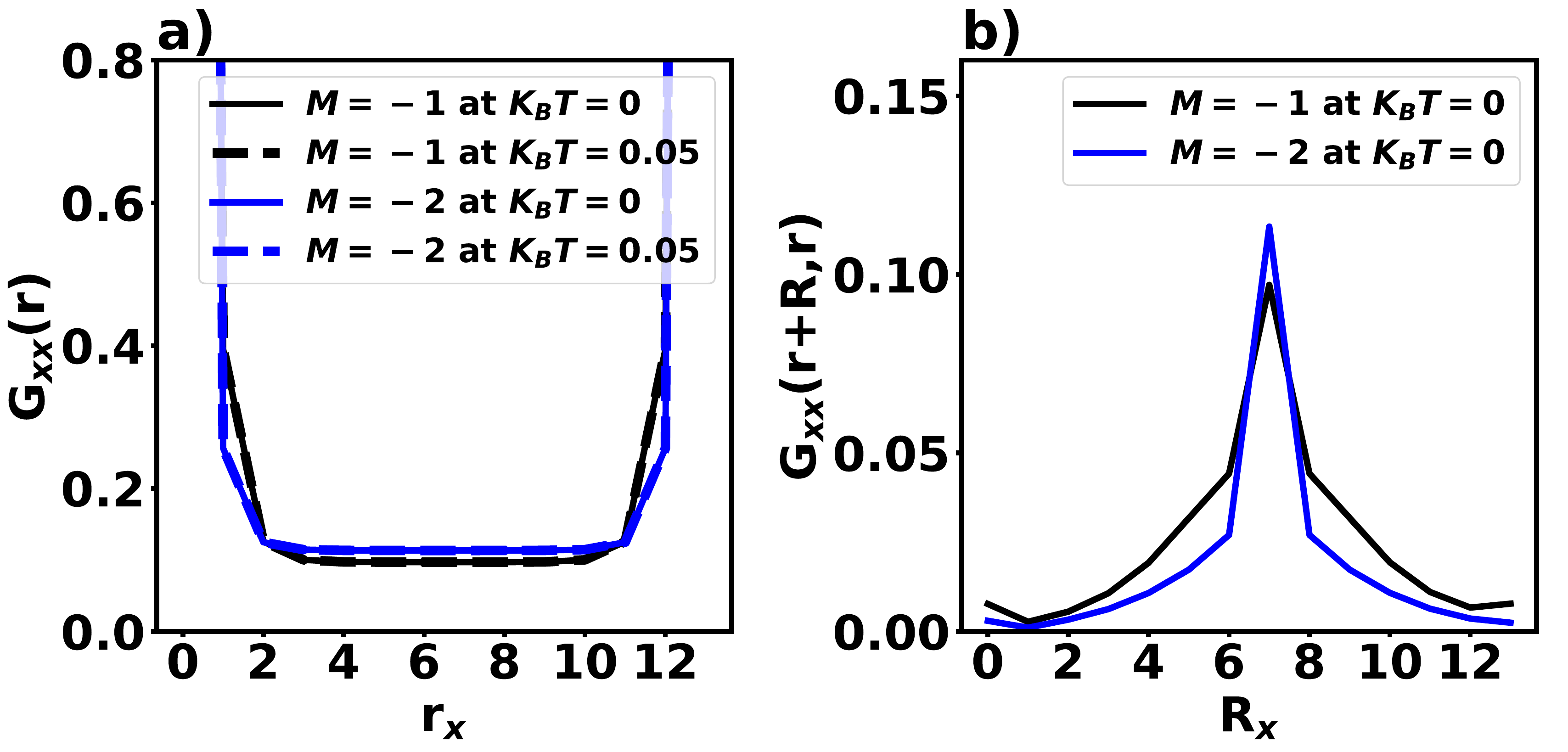}
\caption{Numerical results of the 2D Chern insulator, where we show (a) the fidelity marker ${\cal G}_{xx}(r)$ along ${\hat{\bf x}}$-direction at far from ($M=-2$, blue lines) and close to ($M=-1$, black lines) the critical point $M_{c}=0$, and at zero and finite temperatures, and (b) the spatial profile of the nonlocal fidelity marker.} 
\label{fig:2DChern_results}
\end{center}
\end{figure}

\subsubsection{2D Chern insulator}

We proceed to use the lattice model of Chern insulator as a concrete example for 2D TIs, which is described by the spinless basis $\left(c_{{\bf k}s},\;c_{{\bf k}p}\right)^{T}$ and the momentum space Hamiltonian\cite{Qi11,Bernevig13}
\begin{eqnarray}
H({\bf k})&=&A\sin k_{x}\sigma^{x}+A\sin k_{y}\sigma^{y}
\nonumber \\
&+&\left(M+4B-2B\cos k_{x}-2B\cos k_{y}\right)\sigma^{z},
\label{Hamiltonian_2DclassA_kspace}
\end{eqnarray} 
The corresponding lattice model in real space is\cite{Chen20_absence_edge_current}
\begin{eqnarray}
&&H=\sum_{i}t\left\{-ic_{is}^{\dag}c_{i+ap}
+ic_{i+as}^{\dag}c_{ip}+h.c.\right\}
\nonumber \\
&&+\sum_{i}t\left\{-c_{is}^{\dag}c_{i+bp}+c_{i+bs}^{\dag}c_{ip}+h.c.\right\}
\nonumber \\
&&+\sum_{i\delta}t'\left\{-c_{is}^{\dag}c_{i+\delta s}+c_{ip}^{\dag}c_{i+\delta p}+h.c.\right\}
\nonumber \\
&&+\sum_{i}\left(M+4t'\right)\left\{c_{is}^{\dag}c_{is}
-c_{ip}^{\dag}c_{ip}\right\}-\sum_{iI}\mu\,c_{iI}^{\dag}c_{iI},
\label{Hamiltonian_2DclassA}
\end{eqnarray} 
where we set the parameters to be $t=A/2=1$ and $t'=B=1$, $\sigma=\left\{s,p\right\}$ are the orbitals, and $\delta=\left\{a,b\right\}$ are the lattice constants. We will focus on the $M_{c}=0$ critical point where the bulk gap closes at ${\bf k}=(0,0)$, since the critical behavior near other critical points are similar, and plot the marker along the $(1,0)$ crystalline direction. From Fig.~\ref{fig:2DChern_results} (a) one sees that the marker saturates to a constant value deep inside the bulk that agrees with the momentum integration in Eq.~(\ref{Gmunu_definition}), and the value slightly decreases as temperature increases owing to the thermal broadening. Interestingly, the fidelity marker at $M=-2$ is higher than that at $M=-1$, even though the later is closer to the critical point $M_{c}=0$, which seems to contradict the continuous model in Sec.~\ref{sec:continuous_models}. We find that this is because the real lattice models can exhibit a more sophisticate momentum profile of the quantum metric beyond the Lorentzian shape of Eq.~(\ref{gmumu_Lorentzian}), hence its momentum integration may not monotonically increase as $M\rightarrow M_{c}$. The divergence of fidelity number in this model only takes place roughly in the range $-0.2<M<0$, which requires a much larger lattice to capture. Nevertheless, the nonlocal marker in Fig.~\ref{fig:2DChern_results} (b) simulated on a $14\times 14$ lattice already shows a clear decay with distance ${\bf R}$, with a decay length that increases as $M\rightarrow 0$, indicating that the nonlocal marker serves as a faithful indicator for TPTs in this model even at such a small lattice size.

\begin{figure}[ht]
\begin{center}
\includegraphics[clip=true,width=0.99\columnwidth]{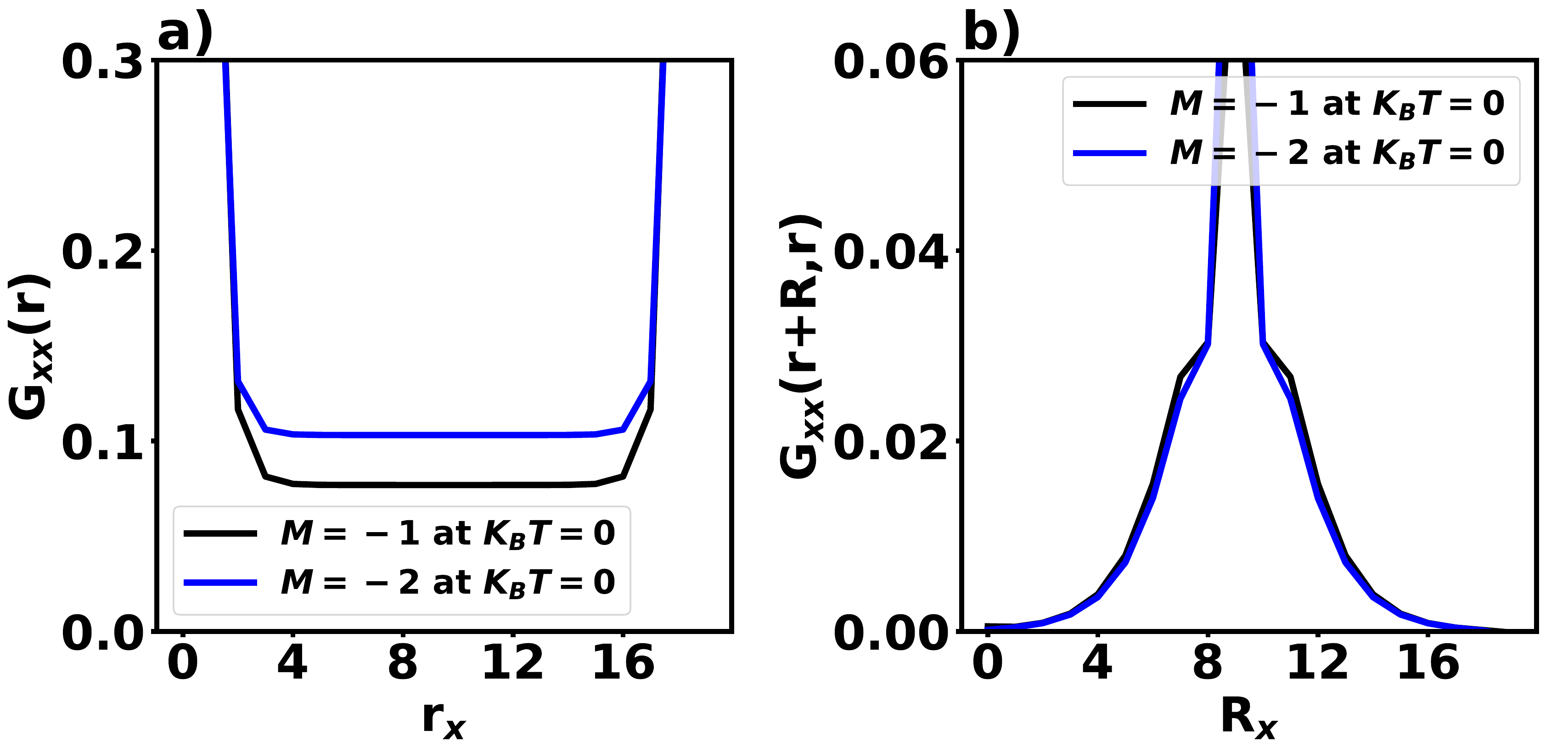}
\caption{Numerical results of the 3D time-reversal symmetric TIs, where we show (a) the fidelity marker ${\cal G}_{xx}({\bf r})$ along ${\hat{\bf x}}$-direction at far from $M=-2$ and close to $M=-1$ the critical point $M_{c}=0$ at zero temperature. (b) The spatial profile of the nonlocal fidelity marker ${\cal G}_{xx}({\bf r+R,r})$ along the same direction.} 
\label{fig:3DTI_results}
\end{center}
\end{figure}

\subsubsection{3D time-reversal symmetric TIs}

For 3D TIs, we consider a class AII model that preserves time-reversal symmetry and is relevant to materials such as Bi$_{2}$Se$_{3}$ and Bi$_{2}$Te$_{3}$, described by the $\Gamma$-matrices and the spinor\cite{Zhang09,Liu10}
\begin{eqnarray}
&&\Gamma^{\ell}=\left\{\sigma^{x}\otimes\tau^{x},\sigma^{y}\otimes\tau^{x},\sigma^{z}\otimes\tau^{x},
I_{\sigma}\otimes\tau^{y},I_{\sigma}\otimes\tau^{z}\right\},
\nonumber \\
&&\psi_{\bf k}=\left(c_{{\bf k}P1_{-}^{+}\uparrow},c_{{\bf k}P2_{+}^{-}\uparrow},c_{{\bf k}P1_{-}^{+}\downarrow},c_{{\bf k}P2_{+}^{-}\downarrow}\right)^{T}
\nonumber \\
&&\equiv\left(c_{{\bf k}s\uparrow},c_{{\bf k}p\uparrow},c_{{\bf k}s\downarrow},c_{{\bf k}p\downarrow}\right)^{T},
\end{eqnarray} 
where we use $s$ and $p$ to label the $P1_{-}^{+}$ and $P2_{+}^{-}$ orbitals in real materials. Keeping only lowest order terms, the low energy Hamiltonian is given by
\begin{eqnarray}
\hat{H}&=&\left(M+M_{1}k_{z}^{2}+M_{2}k_{x}^{2}+M_{2}k_{y}^{2}\right)\Gamma^{5}
\nonumber \\
&+&B_{0}\Gamma^{4}k_{z}
+A_{0}\left(\Gamma^{1}k_{y}-\Gamma^{2}k_{x}\right).
\label{3D_TI_H0_H1}
\end{eqnarray}
The regularization on the whole BZ and a Fourier transform to real space yields the lattice model\cite{Chen20_absence_edge_current}
\begin{eqnarray}
&&H=-\sum_{iI\sigma}\mu c_{iI\sigma}^{\dag}c_{iI\sigma}+\sum_{i\in TI,\sigma}\tilde{M}\left\{c_{is\sigma}^{\dag}c_{is\sigma}-c_{ip\sigma}^{\dag}c_{ip\sigma}\right\}
\nonumber \\
&&+\sum_{i\in TI,I}t_{\parallel}\left\{c_{iI\uparrow}^{\dag}c_{i+a\overline{I}\downarrow}
-c_{i+aI\uparrow}^{\dag}c_{i\overline{I}\downarrow}+h.c.\right\}
\nonumber \\
&&+\sum_{i\in TI,I}t_{\parallel}\left\{-ic_{iI\uparrow}^{\dag}c_{i+b\overline{I}\downarrow}
+ic_{i+bI\uparrow}^{\dag}c_{i\overline{I}\downarrow}+h.c.\right\}
\nonumber \\
&&+\sum_{i\in TI,\sigma}t_{\perp}\left\{-c_{is\sigma}^{\dag}c_{i+cp\sigma}+c_{i+cs\sigma}^{\dag}c_{ip\sigma}+h.c.\right\}
\nonumber \\
&&-\sum_{i\in TI,\sigma}M_{1}\left\{c_{is\sigma}^{\dag}c_{i+cs\sigma}-c_{ip\sigma}^{\dag}c_{i+cp\sigma}+h.c.\right\}
\nonumber \\
&&-\sum_{i\in TI,\delta,\sigma}M_{2}\left\{c_{is\sigma}^{\dag}c_{i+\delta s\sigma}-c_{ip\sigma}^{\dag}c_{i+\delta p\sigma}+h.c.\right\},
\label{3DTIFMM_Hamiltonian}
\end{eqnarray}
where $\tilde{M}=M+2M_{1}+4M_{2}$, $t_{\parallel}=A_{0}/2$, $t_{\perp}=B_{0}/2$, $\left\{I,\overline{I}\right\}=\left\{p,s\right\}$, $\sigma=\left\{\uparrow,\downarrow\right\}$ is the spin index, and $\delta=\left\{a,b,c\right\}$ denotes the lattice constants. We use the parameters $\mu=0.5$ and $t_{\parallel}=t_{\perp}=M_{1}=M_{2}=1$, and two values of $M=-2$ and $M=-1$ that are far from and close to the critical point $M_{c}=0$, respectively.

Figure \ref{fig:3DTI_results} (a) shows the numerical results of ${\cal G}_{xx}({\bf r})$ simulated on a $L_{x}\times L_{y}\times L_{z}=20\times 8\times 8$ lattice plotted along the elongated ${\hat {\bf x}}$ direction. Owing to the significant computational cost at finite temperature, we will focus on the zero temperature limit of this model. The results in Fig.~\ref{fig:3DTI_results} (a) indicates that deep inside the bulk, ${\cal G}_{xx}({\bf r})$ saturates to a constant value that agrees with momentum-integration of the quantum metric $g_{xx}({\bf k})$ in Eq.~(\ref{Gmunu_definition}), and again the value does not monotonically increase at the system approaches the critical point $M\rightarrow 0$ owing to the complicated momentum profile of $g_{xx}({\bf k})$ in the lattice model. Nevertheless, the nonlocal marker ${\cal G}_{xx}({\bf r+R,r})$ shown in Fig.~\ref{fig:3DTI_results} (b) becomes more long-ranged as $M\rightarrow 0$, and hence the nonlocal marker still serves as a faithful index to identify the TPTs in this lattice model.


\begin{figure}[ht]
\begin{center}
\includegraphics[clip=true,width=0.99\columnwidth]{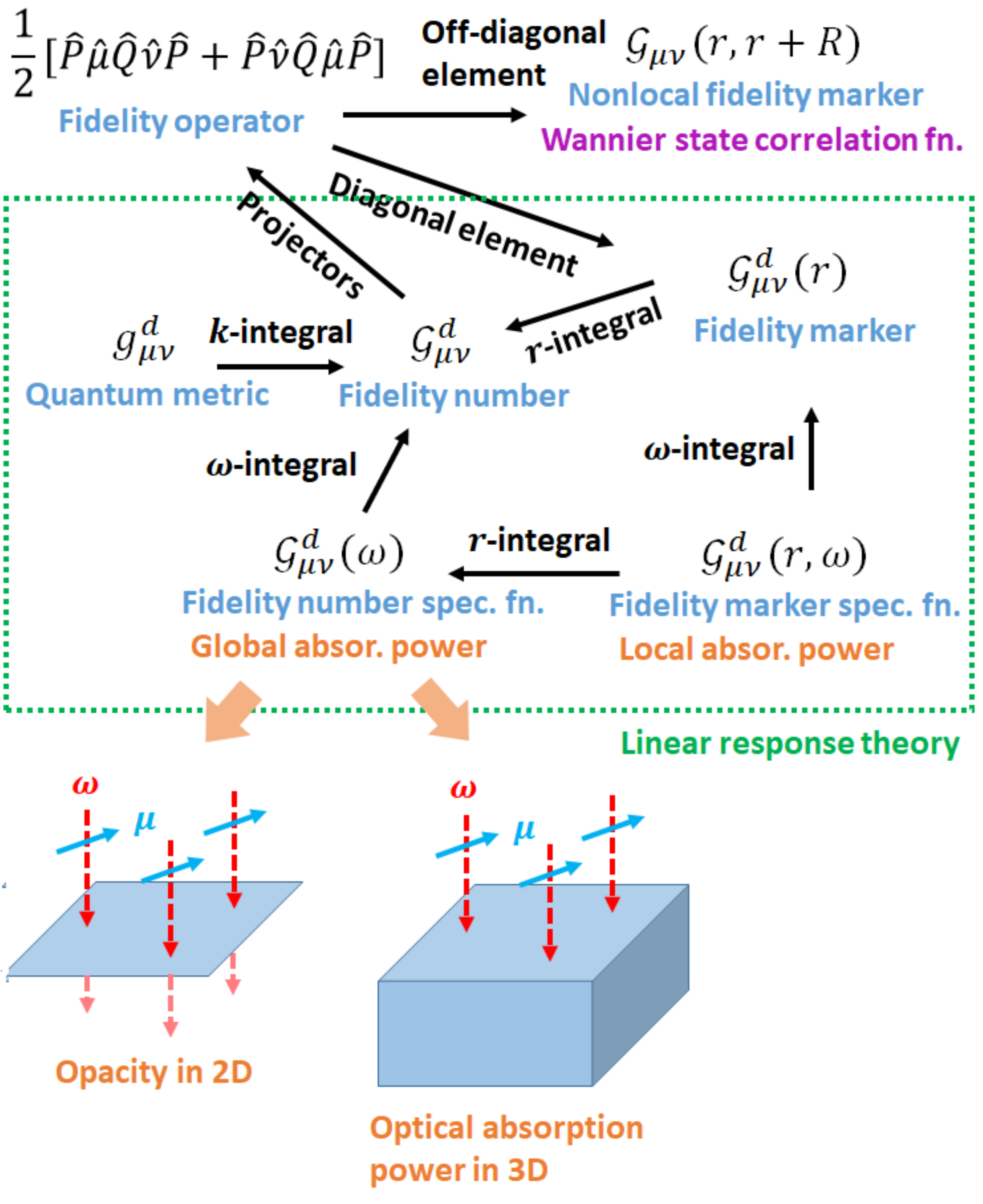}
\caption{Schematics of the linear response theory, measurement protocols, and Wannier function interpretation of various quantities related to the fidelity number and marker. The terminology is given by the blue text, and the orange text indicates the corresponding physical quantity in the proposed optical absorption experiment using linearly polarized light shown in the two figures, and the black arrows indicate how to derive one quantity from another. We use the abbreviations absor.$=$ absorption, spec.$=$ spectral, and fn. $=$ function. } 
\label{fig:summary_diagram_fidelity}
\end{center}
\end{figure}

\section{Conclusions}

In summary, we investigate the fidelity number ${\cal G}_{\mu\nu}$ defined from momentum integration of the quantum metric $g_{\mu\nu}$ of the fully antisymmetric valence band state, which quantifies the average distance between neighboring valence band states in the BZ torus. Using a linear response theory, we generalize the fidelity number to finite temperature denoted by ${\cal G}_{\mu\nu}^{d}$, and suggest that they correspond to the optical absorption power against linearly polarized light, whose frequency-dependence is described by a spectral function ${\cal G}_{\mu\nu}^{d}(\omega)$. Especially for 2D systems, this spectral function can be simply extracted from the frequency-dependence of the opacity measured in units of fine structure constant. We then show that the fidelity number can be mapped to real space as a fidelity marker ${\cal G}_{\mu\nu}^{d}({\bf r})$ defined locally on every unit cell, whose spectral function ${\cal G}_{\mu\nu}^{d}({\bf r},\omega)$ corresponds to the local heating rate that may be measured by atomic scale thermal probes, offering a possibility to measure the gauge-invariant part of the spread of Wannier functions. Moreover, in contrast to the diagonal elements of the fidelity operator that give the local fidelity marker, the off-diagonal elements of the same operator yield a nonlocal fidelity marker ${\cal G}_{\mu\nu}({\bf r+R,r})$ that is equivalent to the Fourier transform of the quantum metric, and represents an overlap between Wannier states. The relation between various quantities introduced in the present work is summarized in Fig.~\ref{fig:summary_diagram_fidelity}.


Particularly for TIs, the quantum metric $g_{\mu\nu}$ diverges at the gap-closing HSPs at the system approaches TPTs, rendering a fidelity number ${\cal G}_{\mu\nu}^{d}$ that diverges at the critical point in 1D and 2D, while approaching a constant in 3D. The predicted fidelity number spectral function ${\cal G}_{\mu\nu}^{d}(\omega)$ also strongly depends on the dimension of the system, and is readily measurable in realistic TIs by optical absorption. The nonlocal marker ${\cal G}_{\mu\nu}({\bf r+R,r})$ is found to decay with a correlation length $\xi$ that diverges near TPTs, with a critical exponent $\nu=1$ that is universal for linear Dirac models in any dimension and symmetry class. We then use prototype lattice models of TIs in 1D, 2D, and 3D to investigate the behavior these markers near critical points in realistic materials, suggesting the ubiquity of these markers in characterizing the quantum geometry and quantum criticality in topological materials. 


These properties of fidelity number and marker immediately imply a great number of applications. Firstly, because the marker is locally defined, it may be used to explore the influence of real space inhomogeneity, such as impurities and grain boundaries, on the quantum geometry of the material. Secondly, because the nonlocal fidelity markers is equivalently the Fourier transform of the quantum metric, we postulate that it can detect any quantum phase transitions provided momentum remains a good quantum number. This conjecture is made because quantum metric $g_{\mu\nu}({\bf k})$ is essentially the fidelity susceptibility of the valence band state defined with respect to momentum ${\bf k}$\cite{You07,Zanardi07,Gu08,Yang08,Albuquerque10,Gu10,Carollo20}, which is expected to diverge at quantum phase transitions, causing the decay length of the nonlocal marker to diverge accordingly. Even if the transition is driven by weak interactions, given the recent generalization of quantum metric to interacting systems\cite{Chen22_dressed_Berry_metric}, it may still be possible to define the fidelity marker perturbatively in terms of Green's function and investigate its critical behavior near the transition. All of these intriguing questions await further investigations to clarify, which may help to shed some new insight on quantum geometry and quantum phase transitions.  

\begin{acknowledgments}

The authors acknowledge stimulating discussions with A. Marrazzo and J. Mitscherling, which help us to clarify various aspects about quantum metric and the Wannier state interpretations, as well as the financial support from the productivity in research fellowship from CNPq. 

\end{acknowledgments}

\appendix 

\section{Comparison with the localization marker \label{apx:comparison_localization_marker}}

We now make a detailed comparison between our fidelity marker and the localization marker proposed in Ref.~\onlinecite{Marrazzo19}. Normalizing the spatial sum of localization marker ${\cal L}_{\mu\nu}$ to the same unit as the fidelity number ${\cal G}_{\mu\nu}$ and use the notation in the present work, the expression is
\begin{eqnarray}
&&{\cal L}_{\mu\nu}=\frac{1}{V}\left(\langle{\hat\mu}{\hat\nu}\rangle-\langle{\hat\mu}\rangle\langle{\hat\nu}\rangle\right)
=-\frac{\hbar^{D-2}}{a^{D}N}{\rm Tr}\left\{{\hat P}\left[{\hat\mu},{\hat P}\right]\left[{\hat\nu},{\hat P}\right]\right\}
\nonumber \\
&&=\frac{\hbar^{D-2}}{a^{D}N}{\rm Tr}\left[{\hat P}{\hat\mu}{\hat Q}{\hat \nu}{\hat P}\right]
=\frac{1}{N}\sum_{\bf r}{\cal{\overline L}}_{\mu\nu}({\bf r}), 
\label{localization_number_definition}
\end{eqnarray}
where $\langle...\rangle$ denotes the average over Wannier states, which defines the localization marker ${\cal{\overline L}}_{\mu\nu}({\bf r})$. If we call ${\cal{\hat L}}_{\mu\nu}\equiv{\hat P}{\hat\mu}{\hat Q}{\hat \nu}{\hat P}$ the operator that renders the localization marker, then a direct comparison with our fidelity operator ${\hat {\cal G}_{\mu\nu}}\equiv\left[{\hat P}{\hat\mu}{\hat Q}{\hat \nu}{\hat P}+{\hat P}{\hat\nu}{\hat Q}{\hat \mu}{\hat P}\right]/2$ in Eq.~(\ref{fidelity_number_definition}) immediately suggests that the diagonal components of them are equal ${\hat {\cal G}_{\mu\mu}}={\hat {\cal L}_{\mu\mu}}$, and the off-diagonal components are related by ${\hat {\cal G}_{\mu\nu}}|_{\mu\neq\nu}=({\hat {\cal L}_{\mu\nu}}+{\hat {\cal L}_{\nu\mu}})/2$. As a result, the diagonal components of the fidelity marker is exactly equal to that of the localization marker. Moreover, shall one define a nonlocal localization marker by ${\cal{\overline L}}_{\mu\nu}({\bf r+R,r})=\langle{\bf r+R}|{\cal{\hat L}}_{\mu\nu}|{\bf r}\rangle$, then it is clear that the diagonal components of it will be equal to our nonlocal fidelity marker too. However, the off-diagonal components are not the same, since our fidelity marker is symmetrized. In summary,
\begin{eqnarray}
&&{\cal G}_{\mu\mu}({\bf r})={\cal{\overline L}}_{\mu\mu}({\bf r}),
\nonumber \\
&&{\cal G}_{\mu\mu}({\bf r+R,r})={\cal{\overline L}}_{\mu\mu}({\bf r+R,r}),
\nonumber \\
&&{\cal G}_{\mu\nu}|_{\mu\neq\nu}({\bf r})=\left[{\overline {\cal L}_{\mu\nu}}({\bf r})+{\overline {\cal L}_{\nu\mu}}({\bf r})\right]/2,
\nonumber \\
&&{\cal G}_{\mu\nu}|_{\mu\neq\nu}({\bf r+R,r})=\left[{\overline {\cal L}_{\mu\nu}}({\bf r+R,r})+{\overline {\cal L}_{\nu\mu}}({\bf r+R,r})\right]/2.
\nonumber \\
\end{eqnarray} 
We emphasize that our definition of the marker stems from the consideration of experimental measurability, especially its link to the optical absorption power in 3D and opacity in 2D, as explained in Sec.~\ref{sec:linear_response}. In particular, the off-diagonal element ${\cal G}_{\mu\nu}|_{\mu\neq\nu}$ defined in our way corresponds to the difference in absorption power or opacity of two specific polarizations, as elaborated in Eqs.~(\ref{DeltaWa_Gxy}) and (\ref{opacity_Gxy}), which is readily measurable. Moreover, the symmetrized form of ${\hat {\cal G}}_{\mu\nu}$ is also a natural consequence of the valence state quantum metric in Eq.~(\ref{gmunu_T0}) that is symmetric in the two indices $\mu$ and $\nu$. Finally, through comparing with these previous works, one sees that our experimental proposal can directly measure the so-called gauge invariant part of the spread of Wannier functions given by the first line of Eq.~(\ref{localization_number_definition}) and usually denoted by $\Omega_{I}$\cite{Marzari97,Souza00}, which may help to compare the first principle calculation with experimental results. 


\bibliography{Literatur}

\end{document}